\documentclass[superscriptaddress,aps,prd,twocolumn,showpacs,nofootinbib,longbibliography, superscriptaddress, showkeys]{revtex4-2}

\usepackage{amsmath,amssymb,amsthm}
\usepackage{calligra,lmodern}
\usepackage{amstext}
\usepackage{easybmat}
\usepackage[colorlinks=true,citecolor=blue,urlcolor=blue, linkcolor= magenta]{hyperref}
\usepackage[pdftex]{graphicx}
\usepackage{times, txfonts}
\usepackage{braket}
\usepackage{color}
\usepackage{ragged2e}
\usepackage{booktabs}
\usepackage{epstopdf}
\usepackage{gensymb}
\usepackage{appendix}
\usepackage{natbib}
\usepackage{float}
\usepackage{caption}
\usepackage{subcaption}
\usepackage{float}
\usepackage{textcase}
\newcommand{\be}{\begin{equation}}
	\newcommand{\ee}{\end{equation}}
\newcommand{\ba}{\begin{eqnarray}}
	\newcommand{\ea}{\end{eqnarray}}

\usepackage{tikz,xcolor,hyperref}

\definecolor{lime}{HTML}{A6CE39}
\DeclareRobustCommand{\orcidicon}{\hspace{-4pt}
	\begin{tikzpicture}
		\draw[lime, fill=lime] (0,0) 
		circle [radius=0.16] 
		node[white] {\hspace{0.1mm}{\fontfamily{qag}\selectfont \tiny ID}};
		\draw[white, fill=white] (-0.07,0.1) 
		circle [radius=0.01];
	\end{tikzpicture}
	\hspace{-3.2mm}
}

\foreach \x in {A, ..., Z}{\expandafter\xdef\csname 
	orcid\x\endcsname{\noexpand\href{https://orcid.org/\csname orcidauthor\x\endcsname}
		{\noexpand\orcidicon}}
}

\begin{document}

\title{ Hawking Radiation in $f(\mathcal{R})$ Gravity: Survival of lighter black holes}
\author{Mriganka Dutta\orcidA{}}\email{mrigankad@iisc.ac.in}
\affiliation{Department of Physics, Indian Institute of Science, Bangalore 560012, India}
\author{Panchajanya Dey}\email{panchajanyad@iisc.ac.in}
\affiliation{Department of Physics, Indian Institute of Science, Bangalore 560012, India}

\author{Banibrata Mukhopadhyay\orcidB{}}\email{bm@iisc.ac.in}
\affiliation{Department of Physics, Indian Institute of Science, Bangalore 560012, India}

\begin{abstract}
     Einstein's theory of general relativity (GR) has been remarkably successful in describing gravitational phenomena. However, several open questions in modern cosmology and astrophysics (e.g. inflation, dark energy) suggest the need for extensions or modifications to this framework. Modified gravity (MGR) theories, including scalar-tensor models and higher-dimensional approaches, attempt to address these gaps while maintaining consistency with established experimental tests.
    This work investigates Hawking radiation within an $f(\mathcal{R})$ gravity theory, with $\mathcal{R}$ being scalar curvature, focusing on its implications for primordial black holes (PBHs) as potential dark matter (DM) candidates. Our analysis reveals that PBHs evaporate slowly in MGR compared to GR predictions. Specifically, we find that non-rotating black holes with masses $\sim 5 \times 10^{13}$ g or lower would have survived by the present epoch, depending on the MGR parameter—a mass threshold approximately at least ten times smaller than in GR. This retarded evaporation timeline imposes relaxed new constraints on the viability of PBHs as DM constituents, thereby reshaping the landscape of possible solutions to the DM problem. This motivates further investigation into alternative gravitational theories and their cosmological consequences.
\end{abstract}    
\maketitle

\section{Introduction}

The theory of Einstein's general relativity (GR) is one of the most successful classical theories in modern times, established in 1915. This non-linear theory revolutionized the idea of spacetime interaction with matter, radiation, and vise versa. There are multiple aspects of modern astronomy where the consequences of GR have been observed, such as Mercury's perihelion precession \cite{Kraniotis:2003ig, Park:2017zgd}, gravitational lensing \cite{Wambsganss:1998gg}, gravitational waves \cite{LIGOScientific:2016aoc, LIGOScientific:2017vwq}, etc. However, many of these phenomena have been predicted considering weak field approximations of GR. It is still uncertain whether this theory is absolute at high spacetime curvature regions. 

The presence of dark matter (DM), which constitutes about 26.4\% of the observable universe, remains one of the most intriguing problems in our current understanding of the cosmos. Its fundamental nature is still unknown; the only established property is that it interacts predominantly through gravity. Not only DM, but certain aspects of early-universe dynamics, such as cosmic inflation, also remain unresolved problems. Consequently, there have been multiple attempts to capture these effects through alternative and modified theories of gravity (MGRs), for example, scalar-tensor theory \cite{Brans:1961sx,Quiros:2019ktw}, $f(\mathcal{R})$ gravity \cite{Sotiriou:2008rp}, with $\mathcal{R}$ being scalar curvature, etc.

The landscape of MGRs was fundamentally transformed by Starobinsky's pioneering work in 1980 \cite{STAROBINSKY198099}, which introduced the first viable $f(\mathcal{R})$ gravity model with the action
\begin{align}
f(\mathcal{R}) = \mathcal{R} + \alpha \mathcal{R}^2,
\label{1.1}
\end{align}
where $\alpha$ is the MGR parameter.
This theory emerged as a compelling alternative to scalar-field inflation, demonstrating that higher-order curvature terms could naturally drive exponential expansion of the early universe. Starobinsky's framework established the $f(\mathcal{R})$ gravity as a serious candidate for addressing both early- and late-time cosmic acceleration while maintaining better behavior than GR. Nevertheless, this does not explain the existence of DM
(though see \cite{Milgrom:1983pn,Bekenstein:1984tv,Milgrom:1992hr}).

Over the years, it is believed that primordial black holes (PBHs) are one of the primary candidates of DM \cite{Bertone:2016nfn, Cirelli:2024ssz}. It, however, has been seen that all PBHs can not serve as the DM present today. Generally, from different observations we are able to put constraints on fraction of PBHs ($f_{PBH}$)
existing as DM \cite{villanueva2021brief}. However, the mass of a PBH can change via Hawking radiation.

The concept of Hawking radiation was introduced in 1975, applying quantum field theory in GR background \cite{Hawking1975, Hawking_1976}.
The quantum mechanical emission of radiation from black holes (BHs) is a semi-classical phenomenon. In this process, pairs of positive and negative energy modes are generated in the vicinity of the horizon of a BH and can get causally separated by the horizon of the BH, leading to change of different BH properties \cite{Hawking1975, Hawking_1976, Bekenstein_1973}. The positive energy modes go out to spatial infinity via tunneling, producing a net outward energy flux \cite{Parikh:1999mf}. The negative energy modes go into BH, decreasing its energy hence it evaporates. 

It turns out that the radiation rate for different energy modes is dependent on the surface gravity of the BH. The apparent temperature related to this radiation mode is given by \cite{Bekenstein_1973}

\begin{equation}
T_H = \frac{\kappa}{2\pi},
\label{1.2}
\end{equation}
where $\kappa$ is the surface gravity. For a non-rotating BH of mass $M$ in GR, the effective temperature is given by
\begin{equation}
    T_H = \frac{\hbar c^3}{8\pi G M k_B} \approx 6.2\times10^{-8}\left(\frac{M_\odot}{M}\right)\text{K},
    \label{1.3}
\end{equation}
where $M_\odot$ represents solar mass, $\hbar$ the reduced Planck constant, $c$ the speed of light, $G$ the Newton's gravitation constant, and $k_B$ the Boltzmann constant. For a non-rotating BH of initial mass $M$, the timescale for complete evaporation is given by

\begin{equation}
\tau \approx \frac{5120\pi G^2 M^3}{\hbar c^4} \sim 2.1\times10^{64}\left(\frac{M}{M_\odot}\right)^3\text{Yrs}.
\label{1.4}
\end{equation}

Hence, for a stellar mass BH, the temperature is extremely low $\sim 10^{-8}$ K, and it would take much longer than the age of the universe to evaporate through Hawking radiation. Nevertheless, due to the inverse proportionality between the temperature and mass, the less massive a BH is, the faster is its rate of evaporation. This leads to the fact that all non-spinning BHs that are formed in the early universe with an initial mass of $\lesssim 10^{14}$ g, should have been evaporated completely by today \cite{Page:1976df}. However, this result strictly follows GR. 

In the present work, we consider an asymptotically flat vacuum solution of the $f(\mathcal{R})$ gravity model \cite{Kalita:2019xjq}, i.e. its BH solution reduces to GR far away from a compact object but deviates from GR in its vicinity. From our model, we show that it is possible for a BH to survive longer compared to GR-based estimates. This suggests that PBHs, with masses even below $ \sim 10^{14}\,\mathrm{g}$, formed in the early universe, can persist and exist as DM.

The plan of the paper is the following. In the next section, we introduce the $f(\mathcal{R})$ gravity under consideration, proposed earlier \cite{Kalita:2019xjq} and some of its basic properties for the purpose of the work. Section \ref{formalism} describes the formalism to obtain BH evaporation rate due to emissions of various particles. Afterwards, section \ref{grey} calculates the Page factor by means of effective potentials for emission rates for various spins of the fields, and section 
\ref{mass-time} computes the survival mass of PBHs today. Further, section \ref{Solar System Test} verifies the MGR and its parameter with solar system test. In section \ref{mb}, we compare the effect of MGR with memory burden effect, and finally section \ref{Conclusion} ends with  conclusions.

\section{Modified Gravity Theory}

The general $f(\mathcal{R})$ gravity framework extends Einstein-Hilbert action by replacing the scalar curvature (Ricci scalar) $\mathcal{R}$ with a nonlinear function $f(\mathcal{R})$, yielding the action
\begin{equation}
    S = \int \left[\frac{c^4}{16\pi G}f(R)+\mathcal{L}_M\right]\sqrt{-g}d^4 x.
    \label{2.1}
\end{equation}
Variation with respect to the metric produces the following field equation
\begin{equation}
f'(\mathcal{R})\mathcal{R}_{\mu\nu} - \frac{1}{2}f(\mathcal{R})g_{\mu\nu} - \left[\nabla_\mu\nabla_\nu - g_{\mu\nu}\Box\right]f'(\mathcal{R}) = \kappa T_{\mu\nu}.
\label{2.2}
\end{equation}
MGR under consideration will have \cite{Kalita:2019xjq} 
\begin{equation}
 F(\mathcal{R}) = \mathrm{d}f(\mathcal{R})/\mathrm{d}\mathcal{R}=1 + B/r,   
\label{FR}
\end{equation}
such that the action reduces to Einstein-Hilbert action at $r\rightarrow \infty$, where $B$ is the MGR parameter. In this theory $ B>0$ gives repulsive nature of gravity outside the compact object, which is unphysical, hence we only consider $ B\leq 0$ and $B=0$ gives GR.

\begin{figure}
    \centering
    \includegraphics[width=0.9\linewidth]{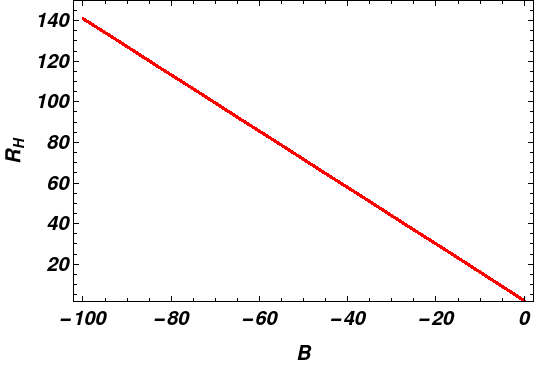}
    \caption{Variation of event horizon radius with $B$.}
    \label{fig1}
\end{figure}

 Assuming a spherically symmetric vacuum solution of MGR, the metric is chosen as
\begin{equation}
    g_{\mu\nu} = diag\biggl(-s(r),p(r),r^2,r^2sin^2\theta\biggr)
    \label{2.3},
\end{equation}
where
\begin{multline}
 s(r) = 1 - \frac{2}{r} -\frac{B(-6 + B)}{2r^2} + \frac{B^2(-66 + 13B)}{20r^3} \\- \frac{B^3(-156 + 31B)}{48r^4} + \frac{3B^4(-57 + 11B)}{56r^5} \\
    -\frac{B^5(-360 + 67B)}{128r^6} + \mathcal{O}(r^{-6})
    \label{2.4}
\end{multline}
and $p(r) = X(r)/s(r)$, with $X(r)$ is given by
\begin{equation}
    X(r) = \frac{16r^4}{(B+2r)^4},
    \label{2.5}
\end{equation}
as already was established earlier \cite{Kalita:2019xjq}.
Throughout the paper we consider $G=M=c=1$, unless stated otherwise. The radius of event horizon ($R_H$) for this BH metric has been calculated from Eq.\,(\ref{2.4}) by considering till $20^{\mathrm{th}}$ order of $1/r$.
Fig.\,\ref{fig1} shows that $R_H$ increases with increasing $|B|$. This implies that MGR allows a larger event horizon compared to GR.

\begin{figure}[H]
    \centering
    \includegraphics[width=0.85\linewidth]{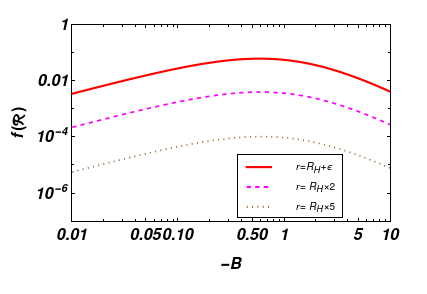}
    \caption{Variation of $f(\mathcal{R})$ with $B$ at different $r$.}
    \label{fig9}
\end{figure}

Using Eqs.~(\ref{2.4}) and (\ref{2.5}), it is easy to calculate $\mathcal{R}$ as a function of $r$ and $B$ following GR norms. This $\mathcal{R}$ turns out to be non-zero outside the BH horizon. Then the straightforward calculation, by integrating $F(\mathcal{R})$ from Eq. (\ref{FR}), gives
\begin{multline}
    f(\mathcal{R})=\frac{3 (B-2) B}{r^4}+\frac{3 (7 B-4) B^2}{10 r^5}+\frac{(11 B-42) B^3}{20 r^6}+\\\frac{(552-101 B) B^4}{280 r^7}+\frac{(155 B-912) B^5}{448 r^8}+...
\end{multline}
Fig. \ref{fig9} shows that small and large $|B|$, both lead to smaller $f(\mathcal{R})$, which is not difficult to understand. Small $|B|$ tends the gravity to GR when 
$f(\mathcal{R})\rightarrow \mathcal{R}$, and we know that $\mathcal{R}=0$ for a vacuum solution in GR.
For a large $|B|$, the size of $R_H$ is large, leading the BH surface to be far away from BH singularity, which further weakens the gravity and the effect of MGR goes off as it scales as $1/r$ as in Eq. (\ref{FR}). At an intermediate $|B|$, with the increase of 
$|B|$, MGR effect increases leading to the increased $f(\mathcal{R})$ (deviated from GR) until the effect of increased $R_H$ begins to dominate.
Fig. \ref{fig9} also confirms that the effect of MGR decreases with increasing distance from the BH.

\section{Formalism for a non-rotating black hole evaporation }\label{formalism}
The computation of Hawking radiation is related to the nature of the BH metric and its perturbations by the particles around it. Following Hawking's work, Teukolksy, Page, Chandrasekhar etc. worked out the perturbation equations from the BH metric governing the dynamics of different massless fields. We follow the work by Arbey et.al. \cite{Arbey:2021jif,Arbey:2021yke} to construct the perturbation equations, i.e. the Teukolsky equations, for a general spherically symmetric metric and apply for MGR given by Eq. (\ref{2.3}).

The dynamics of a massless field of spin $s$ in this metric can be represented by a wave-function $\Psi_s(r,\theta,\phi,t)$. Following Teukolsky's master equation for a metric having spherical temporal symmetry, we can take an ansatz
\begin{equation}
    \Psi_s = \Phi_s(r)Y^s_{l,m}(\theta,\phi)e^{-i\omega t}.
    \label{3.2}
\end{equation}
Here, $\omega$ is the energy and $Y^s_{l,m}$ are spin $s$ weighted spherical harmonics for the angular mode $l$, $m$, satisfying the equation

\begin{multline}
    \Bigg(\frac{1}{\sin\theta}\partial_\theta (\sin\theta \, \partial_\theta) + \csc^2\theta \partial_\varphi^2 + \frac{2is\cot\theta}{\sin\theta} \partial_\varphi + s \\  - s^2\cot^2\theta + \lambda_\ell^s \Bigg) ~ Y_{\ell,m}^s = 0,
\label{3.3}
\end{multline}
with the separation constant $\lambda_l^s = l(l+1)-s(s+1)$. 
Now, to study the radial dynamics of the field, it is useful to write the radial part of Teukolsky master equation in the form of a Schr\"odinger like wave equation. This can be done by a transformation to the tortoise-coordinates, and a redefinition of the wavefunction. To transform the equation in the tortoise coordinate system $r^\star$, we solve the differential equation
\begin{equation}
    \frac{dr^\star}{dr} = \sqrt{\frac{p(r)}{s(r)}}.
    \label{3.4}
\end{equation}
The transformation to tortoise coordinate is useful because it maps the event horizon of the BH to $r^\star\to-\infty$. Therefore, the transformed equation of motion describes the motion of the particle between $-\infty$ to $+\infty$, without any singularity in between. The radial wavefunction is also redefined as
\begin{equation}
    Z_s = \Phi_s\sqrt{B_s\sqrt{\frac{p(r)}{s(r)}}},
    \label{3.5}
\end{equation}
where the function $B_s$ takes the forms $\sqrt{s(r)/p(r)}h(r)$, $\sqrt{s(r)/p(r)} s(r)h(r)^2$, $\sqrt{s(r)/p(r)}s(r)^2h(r)^3$, and $\sqrt{h(r)/p(r)}s(r)h(r)$, with $h(r)=r^2$, for spins 0, 1, 2, and 1/2, respectively. Under these transformations, the radial equation reduces to
\begin{equation}
    \partial_{s}^{2}Z_{s} + \left(\omega^{2} - V_{s}\big(r(r^{\ast})\big)\right)Z_{s} = 0 \,,
    \label{3.6}
\end{equation}
where $V_s$ is the effective potential faced by the field of spin $s$. The effective potentials for fields of spins 0, 1, 2, 1/2 take the following form, respectively \cite{Arbey:2021yke}
\begin{align}
\label{3.7}
V_0 &= \nu_0 \frac{s}{h} + \frac{\partial_*^2 \sqrt{h}}{\sqrt{h}}, \nonumber\\
V_1 &= \nu_1 \frac{s}{h}, \nonumber\\
V_2 &= \nu_2 \frac{s}{h} + \frac{(\partial_* h)^2}{2h^2} - \frac{\partial_*^2 \sqrt{h}}{\sqrt{h}}, \nonumber\\
V_{1/2} &= \nu_{1/2} \frac{s}{h} \pm \sqrt{\nu_{1/2}} \, \partial_* \left( \sqrt{\frac{s}{h}} \right),
\end{align}
where $\nu_0 = \nu_1 = l(l+1), \nu_2 = l(l+1)-2$ and $\nu_{1/2} = l(l+1)+1/4$.
Now, to find the tunneling probability of a particle at the horizon, we consider a purely ingoing wave function at the horizon. Therefore, the wavefunction has to satisfy the following form:
\begin{equation}
Z(r^{*}) \mathop{\sim}_{r^{*} \rightarrow -\infty} A_{\mathrm{hor}}^{\mathrm{in}} e^{-i\omega r^{*}} \,,
\label{3.8}
\end{equation}
and, at spatial infinity, it will have both ingoing and outgoing components, so it will be of the form:
\begin{equation}
Z(r^{*}) \mathop{\sim}_{r^{*} \to +\infty} A_{\infty}^{\mathrm{in}} e^{-i\omega r^{*}} + A_{\infty}^{\mathrm{out}} e^{+i\omega r^{*}} .
\label{3.9}
\end{equation}
Solving the wave equation given by Eq. (\ref{3.6}), we obtain the values of the amplitudes $A^{\rm in}_{\rm hor}$ and $A^{\rm in}_\infty$, which leads us to the tunneling probability
\begin{equation}
\Gamma_i(\omega,M,x_j) = \left|\frac{A_{\mathrm{hor}}^{in}}{A_{\infty}^{\mathrm{in}}}\right|^2 .
\label{3.10}
\end{equation}
 Since the radial Schr\"odinger like equation was obtained after variable separation, with angular modes $l$ and $m$, $\Gamma_i$ depends on spin as well as $(l,m)$. The emission rate for spin $s$ energy modes per unit time $t$ and energy $\omega$, after introducing $M$ in equations, is given by \cite{Arbey:2021yke}
\begin{equation}
Q_{s}=\frac{\mathrm{d}^2 N_{s}}{\mathrm{d}t \, \mathrm{d}\omega} = \sum_{l,m} \frac{1}{2\pi} \frac{\Gamma_{s}(\omega,M,l,m,x_j)}{e^{\omega/T_H} - (-1)^{2s}} \,,
\label{3.11}
\end{equation}
where $T_H$ is the Hawking temperature of the BH defined earlier at $R_H$ in Eq. (\ref{1.2}) and $x_j$ denotes any additional parameters such as the BH's spin, charge, etc. 
The surface gravity of a non-rotating BH can further be defined by 
\begin{equation}
    \kappa^2 \equiv -\frac{1}{2} \nabla_\mu k_\nu \nabla^\nu k^\mu \Bigg|_{\mathrm{hor}} = \frac{1}{4} \left.\frac{ (s'(r))^2}{s(r)p(r)}\right|_{\mathrm{hor}} \,,
    \label{3.13}
\end{equation}
where $k^\mu$ denotes the timelike Killing vector. 
 From the information of $Q_s$, we can calculate the total energy loss of a BH. This gives us the Page factor $g(M,x_j)$, that governs the equation for a non-rotating BH evolution, defined as \cite{Page:1976df,Page:1976ki,Page:1977um}
\begin{align}
g(M, x_j) &\equiv -M^2 \frac{\mathrm{d}M}{\mathrm{d}t} = M^2 \int_0^{+\infty} \sum_{s}\, \omega\,Q_s (\omega)\,\mathrm{d}\omega.
\label{3.14}
\end{align}
Hence, the BH evolution takes place according to the equation
\begin{align}
 \frac{dM}{dt} &= -\frac{g(M, x_j)}{M^2}.
\label{3.15}
\end{align}

\section{Computation of Page factor}\label{grey}

The BlackHawk code devoloped by Auffinger et al. \cite{Auffinger_2023} provides the tunneling probabilities for Kerr-like BHs, charged BHs, and in some other scenarios. For this work, the Page factors have been calculated using the modified non-rotating BH metric.

\subsection{Surface gravity of a non-rotating BH}
\begin{figure}
    \centering
    \includegraphics[width=0.9\linewidth]{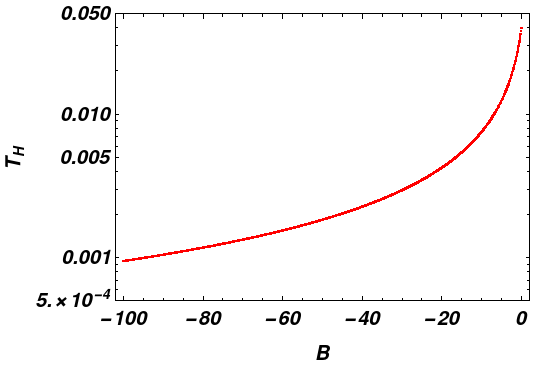}
    \caption{Variation of black hole surface temperature with $B$.}
    \label{fig2}
\end{figure}

 For a static observer at infinity, the surface gravity is the acceleration of a static object near the horizon of a BH. This has been interpreted as the temperature of the BH. The surface gravity of a BH in GR as well as MGR is inversely proportional to its event horizon radius. In MGR we have already seen in Fig.\,\ref{fig1} that the event horizon size of the BH increases with increasing $|B|$. Thus, for a fixed mass of BH, from Eq.\,(\ref{3.13}), the surface gravity decreases as the MGR parameter $B$ becomes more negative. Therefore, from Eq.\,(\ref{1.2}), the BH surface temperature $T_H$ also decreases, as depicted in Fig.\,\ref{fig2}.

\subsection{Effective potential for different fields}
\begin{figure}
    \centering
    \includegraphics[width=0.9\linewidth]{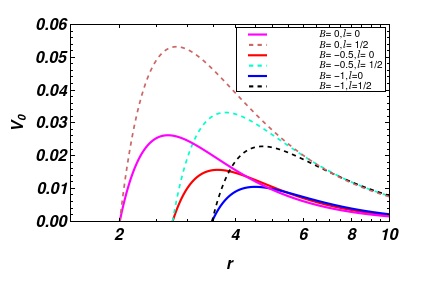}
    \caption{Effective potential for spin 0 particles. }
    \label{fig3}
\end{figure}

\begin{figure}
    \centering
    \includegraphics[width=0.9\linewidth]{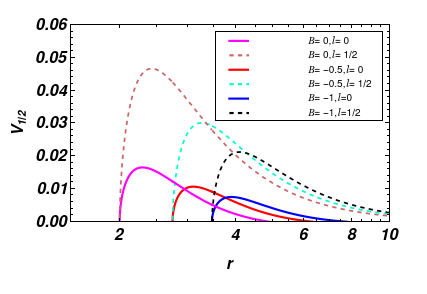}
    \caption{Effective potential for spin $1/2$ particles.}
    \label{fig4}
\end{figure}

 It is clear from Fig.\,\ref{fig1} that $M/R_H$ decreases with the increase of $|B|$ in MGR. Even in GR, e.g. the Schwarzschild metric, we know that decreasing $M$ (or increasing $r$) tends to the Minkowski spacetime, i.e. the effect of BH decreases. Thus, in the same analogy, one can interpret that the potential barrier, originated due to metric perturbation, in the Teukolsky equations will decrease in amplitude with increasing $|B|$. 
 For spins 0 and 1/2 energy modes the potentials given by Eq. (\ref{3.7}) are shown in Figs.\,\ref{fig3} and \ref{fig4}.
\subsection{Emission rates}
\begin{figure}[!htbp]
    \centering
    \includegraphics[width=0.95\linewidth]{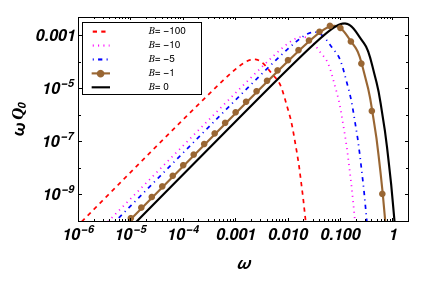}
    \caption{Variation of $\omega Q_0$ with $\omega$ for different $B$. }
    \label{fig5}
\end{figure}

\begin{figure}[!htbp]
    \centering
    \includegraphics[width=0.95\linewidth]{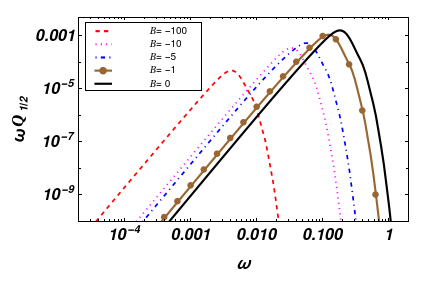}
    \caption{Variation of $\omega Q_{1/2}$ with $\omega$ for different $B$.}
    \label{fig6}
\end{figure}

The rate of emission of spin $s$ energy modes per time $t$ and per energy $\omega$ is $Q_s$. For spin 0 and 1/2 fields, the emission rate of particles as a function of energy is shown in Figs. \ref{fig5} and \ref{fig6}, respectively. 
We see that the rate of emission decreases as we deviate from GR, i.e. increase $|B|$. This is because there are two counteracting factors in the Hawking radiation rate. One of them is the surface area, which increases with increasing $|B|$ in MGR, thereby tending to enhance the emission rate. The other factor is the temperature (surface gravity), which decreases with increasing $|B|$ in MGR, thereby tending to suppress the emission rate. The temperature effect dominates, so the overall rate of emission decreases, with increasing $|B|$ in MGR, relative to GR.

The contributions from spin 1 and spin 2 particles are negligibly small compared to spin 0 and spin 1/2. Therefore, contribution of emission in these cases are neglected.

\subsection{ Page factors}

\begin{figure}[!htbp]
    \centering
    \includegraphics[width=0.85\linewidth]{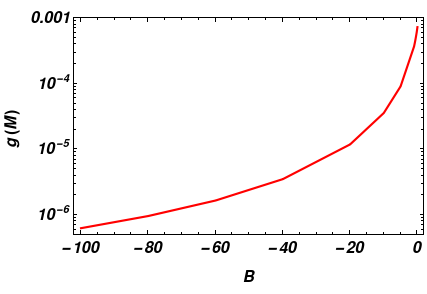}
    \caption{Variation of the Page factor $g(M)$ with $B$.}
    \label{fig7}
\end{figure}

From Eq. (\ref{3.14}), the integrand
$g(M,x_j)$ turns out to be independent of $M$ and it only depends on the model parameters. For the non-rotating BH in MGR, this depends only on the parameter $B$. This is not a surprise, considering that in GR, for a non-rotating BH, also $g(M)$ is constant, which eventually leads to the lifetime of the BH to be proportional to $M^3$. Therefore, in MGR as well, the lifetime of a BH of mass $M$ will be proportional to $M^3$. The variation of Page factor with $B$ is shown in Fig.\,\ref{fig7}.


\section{ mass of PBH that survives in today's universe timescale}\label{mass-time}
As we see in Fig.\,\ref{fig7}, the Page factors decrease in MGR with increasing value of $|B|$. Thus, it is clear from Eq.\,(\ref{3.15}) that the rate of change in BH mass also decreases. Hence, MGR allows even smaller mass PBHs to survive for a longer time scale compared to GR based calculations. 

\begin{figure}[H]
    \centering
    \includegraphics[width=0.85\linewidth]{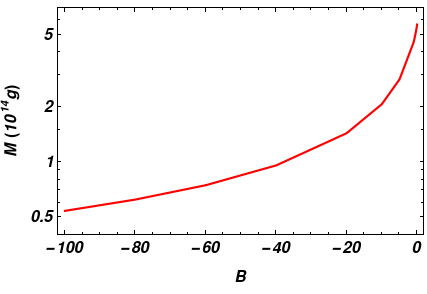}
    \caption{Minimum initial mass of PBH that survives evaporation.}
    \label{fig8}
\end{figure}

 Fig. \ref{fig8} shows the value of the minimum initial mass of a PBH that survives till today for different $B$. It is clear that with increasing $|B|$ smaller and smaller mass survives. In fact, it could be even one order of magnitude or less smaller than what GR predicts, depending on $B$. 

\section{Solar System Test of Modified Gravity}
\label{Solar System Test}
Solar system tests are a way to constrain the parameters of a MGR model by ensuring that the model reduces to GR in asymptotic limits. As discussed by Guo \cite{Guo:2013fda}, an $f(\mathcal{R})$ model of the form $f(\mathcal{R}) = \mathcal{R} + A(\mathcal{R})$ satisfies the solar system test if it obeys the following conditions in the weak field limit:
\begin{equation}
\left|\frac{A(\mathcal{R})}{\mathcal{R}}\right| \ll 1,
\label{6.1}
\end{equation}

\begin{equation}
|A'(\mathcal{R})| \ll 1,
\label{6.2}
\end{equation}

\begin{equation}
\mathcal{R} A''(\mathcal{R}) \ll 1.
\label{6.3}
\end{equation}

In MGR under consideration, the $A(\mathcal{R})$ term takes the form \cite{Kalita:2019xjq} as
\begin{equation}
A(\mathcal{R}) = K_1\mathcal{R}^{5/4}+K_2\mathcal{R}^{3/2}+\dots
\label{6.4}
\end{equation}
where the Ricci scalar $\mathcal{R}$ is of the form
\begin{equation}
\mathcal{R} = \frac{3B(B - 2)}{r^4} - \frac{3B^2(B - 12)}{10 r^5} + \frac{B^3(8B - 51)}{10 r^6}-\dots
\label{6.5}
\end{equation}
and
\begin{equation*}
    K_1 = \frac{12}{5 \times 3^{5/4}} \frac{B}{(B^2 - 2B)^{1/4}}
    \,\, K_2 = \frac{1}{60\sqrt{3}} \frac{(B-12)}{(1-2/B)^{3/2}}. 
\end{equation*}

\begin{figure*}[!htbp]
  \centering
  \begin{subfigure}[b]{0.32\textwidth}
    \centering
    \includegraphics[width=\textwidth]{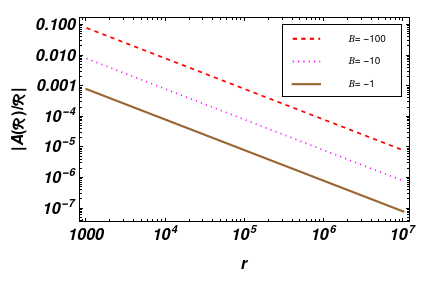}
    \caption{}
    \label{fig10:1}
  \end{subfigure}
  \begin{subfigure}[b]{0.32\textwidth}
    \centering
    \includegraphics[width=\textwidth]{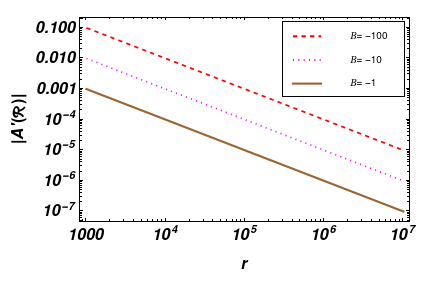}
    \caption{}
    \label{fig10:2}
  \end{subfigure}
  \begin{subfigure}[b]{0.32\textwidth}
    \centering
    \includegraphics[width=\textwidth]{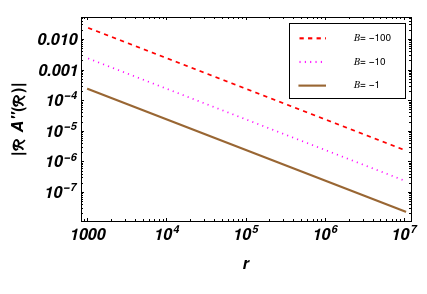}
    \caption{}
    \label{fig10:3}
  \end{subfigure}

  \caption{Conditions for solar system test for different values of $B$.}
  \label{fig10}
\end{figure*}

We see from Fig.\,\ref{fig10} that the solar system test conditions are fully satisfied from low to high values of 
$|B|$. Note that we consider minimum $r$ to be 1000, which is even well below solar radius, similar to small white dwarf radius. Thus, there is no constraint on the value of $B$ just under the consideration of the solar system test.

\section{Memory Burden effect}
\label{mb}
In this section, we compare the effects of MGR with memory burden. A deviation from GR, i.e. MGR, is not the only effect that can suppress the BH evaporation. Recent works of Dvali suggested that the quantum nature of BH radiation can backreact on the BH and slow down the evaporation rate \cite{Dvali:2011aa,Dvali:2020wft}. He proposed that the rate of change in mass of the BH can be suppressed by a factor $S^k(qM_0)$, where $S$ is the entropy, $M_0$ is the initial mass of BH, $q$ is some factor $<1$, and $k$ is a real number which is not constrained. In this scenario, instead of Eq.\,(\ref{3.15}), the evaporation rate follows \cite{Thoss:2024hsr}
\begin{equation}
    \frac{dM}{dt} = -\frac{1}{S^k(qM_0)}\frac{g(M, x_j)}{M^2}.
    \label{7.1}
\end{equation}
According to Dvali, this suppression factor becomes important (at least) after half of BH's mass has evaporated. Thus, for $M> M_0/2$ we have the usual Hawking evaporation rate and after that the evaporation rate is followed by Eq.\,(\ref{7.1}), where we consider $q=1/2$.

\begin{figure}
    \centering
    \includegraphics[width=0.85\linewidth]{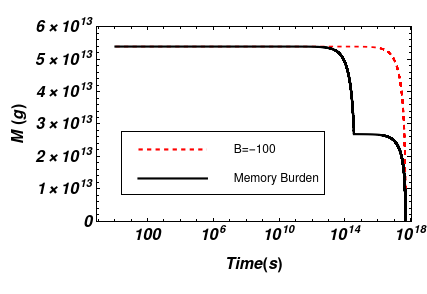}
    \caption{Comparing BH evaporation for $B=-100$ with memory burden effect (with $B=0$) for $k=1.0644\times10^{-1}$.}
    \label{fig11}
\end{figure}

Fig.\,\ref{fig11} shows that the delay of the evaporation time-scale due to MGR, compared to GR, can be mimicked by the memory burden effect on GR. For $B=-100$, a BH of mass $5.4\times10^{13}$ gm just evaporates on the current time-scale. However, following Dvali's proposal, the effect of memory burden sets in once half of the BH mass is evaporated. Hence, a suppression slows down the mass evaporation rate, which survives (or just evaporating) today the same mass as above in GR with $k=0.10644$.

Therefore the evaporation time-scale in MGR for a given $B$ matches with that in GR for an appropriate $k$.
However, for a given $B$, different mass of BH to evaporate completely today requires different $k$ in GR. The reason is that the suppression factor, $S^k(qM_0)$, also depends on the initial mass, $M_0$, of the BH. Therefore, we need different $k$ for different $M_0$ to match with the time-scale of the MGR evaporation case.

\section{Conclusions}
\label{Conclusion}
We have explored Hawking radiation and the survival mass of PBHs in a MGR, which is based on $f(\mathcal{R})$ gravity theory.   
The MGR theory suppresses the rate of evaporation, compared to that in GR, of a non-rotating BH by decreasing the surface gravity. This allows BHs to survive for a longer time. This can affect the current study of constraints on PBHs of the asteroid mass range. A longer survival time means lesser mass PBHs will survive. As a result, the contribution of PBHs to DM will be more in this scenario. The current work with MGR parameter $B = -100$ allows a PBH of mass $5.4\times 10^{13}$ g to survive in today's universe timescale, which is almost ten times lower than GR based calculation.

In the current work, solar system test could not constrain $B$ noticeably and $|B|$ could be very large, still satisfying solar system and even
the features at the surface of white dwarfs.
However, it is still possible to get a strict bound of $B$ from other phenomena, e.g. from the observation of Sgr\,$\mathrm{A}^*$ shadow.

There are many possible MGR theories and we consider only one. Current results suggest to explore Hawking radiation in other MGR theories, if all exclusively decrease evaporation rate, at least the ones which are asymptotically flat which are more suitable astrophysically. They may offer better bound on MGR parameter (equivalent to $B$) as well. Also the possible equivalence of the MGR effects to memory burden can be explored in detail for different models.

\section*{Acknowledgment}
The authors thank Prashant Kocherlakota (CMI) for helpful discussions. MD acknowledges the financial support of the Ministry of Education (MoE) fellowship scheme.
BM thanks the partial support from the project funded by SERB/ANRF,
India, with Ref. No. CRG/2022/003460.


\bibliography{bibliography_md}

\end{document}